\documentclass[showpacs,10pt,twocolumn,prb]{revtex4-1}
\usepackage{amsmath}
\usepackage{amssymb}
\usepackage{graphics}
\usepackage{epsfig}
\usepackage{CJK}
\usepackage{color}
\usepackage{lineno}

\begin{document}

\title{Critical behaviors of half-metallic ferromagnet Co$_{3}$Sn$_{2}$S$_{2}$}
\author{Weinian Yan$^{1}$}
\author{Xiao Zhang$^{1,*}$\thanks{zhangxiaobupt@bupt.edu.cn}}
\author{Qi Shi$^{1}$}
\author{Xiaoyun Yu$^{1}$}
\author{Zhiqing Zhang$^{1}$}
\author{Qi Wang$^{2}$}
\author{Si Li$^{1}$}
\author{Hechang Lei$^{2}$}
\address{$^{1}$State Key Laboratory of Information Photonics and Optical Communications $\&$ School of Science, Beijing University of Posts and Telecommunications, Beijing 100876, China\\
$^{2}$Department of Physics and Beijing Key Laboratory of Opto-electronic Functional Materials $\&$ Micro-nano Devices, Renmin University of China, Beijing 100872, China}

\date{\today}

\begin{abstract}
We have investigated the critical behavior of a shandite-type half-metal ferromagnet Co$_{3}$Sn$_{2}$S$_{2}$. It exhibits a second-order paramagnetic-ferromagnetic phase transition with $T_{C}=$ 174 K. To investigate the nature of the magnetic phase transition, a detailed critical exponent study has been performed. The critical components $\beta$, $\gamma$, and $\delta$ determined using the modified Arrott plot, the Kouvel-Fisher method as well as the critical isotherm analysis are match reasonably well and follow the scaling equation, confirming that the exponents are unambiguous and intrinsic to the material. The determined exponents of Co$_{3}$Sn$_{2}$S$_{2}$ deviates from theoretical estimated short-range universal models. Instead, Co$_{3}$Sn$_{2}$S$_{2}$ exhibits long-range order in the nature of magnetic interaction with the spin decay as $J(r)\sim1/r^{-(d+\sigma)}$ with $\sigma=1.28$.
\end{abstract}

\maketitle

\section{Introduction}

The investigation of phase transition in magnetic materials provides insight into the fundamental interactions and the global properties, such as the space dimensionality, the range of interaction, and the symmetry of the order parameter within a given universality class that the critical exponents and scaling functions are identical for all systems.\cite{SLSondhi,SUNYP,WangPRB,WangJMMM} Most of the theories of critical phenomena mainly study the magnets in which the magnetic moments are localized at lattice sites and interacting with one another through exchange interactions. In contrast, experimental studies on the critical behaviors of itinerant ferromagnets (IFMs) near the ferromagnetism (FM) and paramagnetism (PM) phase transition are confined to only a few systems.\cite{Kouvel,Boxberg,Perumal,Semwal,ABhattacharyya} Itinerant ferromagnetism is based on the band theory of electrons and the magnetic moment arises from the exchange splitting of the band.\cite{Shimizu,Moriya} Although modern theories developed from the Stoner mean field model can explain well various characteristic physical properties of IFMs,\cite{Shimizu,Moriya} such as Curie temperature ($T_{C}$), temperature dependence of spontaneous magnetization, $M(T,0)$ for $T<T_{C}$ and Curie-Weiss behavior of susceptibility, $\chi(T)$ when $T>T_{C}$ etc, they still inadequately explain the critical behaviors of these systems.\cite{Moriya2}

Among IFMs, half-metallic ferromagnets (HMFMs) belong to one of limits. In general, they are metals with a Fermi surface in one spin channel but for the other there is a gap in the spin-polarized density of states (DOS), like a semiconductor or insulator.\cite{Groot,O'Handley} Because of very limited HMFMs reports, such as Heusler alloys,\cite{Groot} CrO$_{2}$,\cite{Coey} and Sr$_{2}$FeMoO$_{6}$,\cite{Kobayashi} the physical properties of HMFMS are not fully understood, let alone their critical behaviors. On the other hand, the HMFMs are also essential for applications related to magnetism and spintronics.\cite{Moodera,Zutic} For example, half-metallic ferromagnetic electrodes can be used as ideal spin injectors and detectors, because they can carry current in only one spin direction. They are also important components for giant magnetoresistance (GMR) and tunneling magnetoresistance (TMR) devices.

Recently, the shandite-type HMFM Co$_{3}$Sn$_{2}$S$_{2}$ with $T_{C}=$ 174 K has attracted much research attention.\cite{Weihrich,WSchnelle,WangQ,LiuEK} It exhibits large intrinsic anomalous Hall effect, originating from the Weyl fermions near the Fermi energy.\cite{WangQ,LiuEK} More importantly, the density-functional theory (DFT) calculations indicate that a ferromagnetic phase transition in this compound is accompanied by the formation of a band gap in spin minority direction.\cite{Weihrich} According to the classification given by Coey,\cite{Coey} Co$_{3}$Sn$_{2}$S$_{2}$ is a type $I_{A}$ HMFM, which is an ideal material for the spintronics devices because of the fully spin polarization at the Fermi energy. Since only very few compounds have been proven to be HMFMs with high $T_{C}$, the detailed experimental study on the physical properties of Co$_{3}$Sn$_{2}$S$_{2}$ is very important. But the investigations on the critical behaviors of Co$_{3}$Sn$_{2}$S$_{2}$ single crystal are still scarce.

In this work, we studied the critical behaviors of Co$_{3}$Sn$_{2}$S$_{2}$ single crystal. For the understanding of the nature of magnetic phase transition, we performed a critical exponent analysis in the vicinity of the FM-PM transition region, where the critical exponents have been obtained reliably by different analytical methods. This is an effective way to clarify the magnetic interactions and properties of Co$_{3}$Sn$_{2}$S$_{2}$.\cite{Fan,Ghosh}

\section{Experimental}

Single crystals of Co$_{3}$Sn$_{2}$S$_{2}$ were grown from Sn flux.\cite{Kassem} X-ray diffraction (XRD) of well-grounded crystals and a single crystal was performed using a Bruker D8 X-ray diffractometer with Cu $K_{\alpha}$ radiation ($\lambda=$ 0.15418 nm) at room temperature. Rietveld refinements of the data were performed with the TOPAS package.\cite{TOPAS} Magnetization measurements were carried out using Quantum Design MPMS3.

\section{Results and discussion}

\begin{figure}[tbp]
\centerline{\includegraphics[scale=0.32]{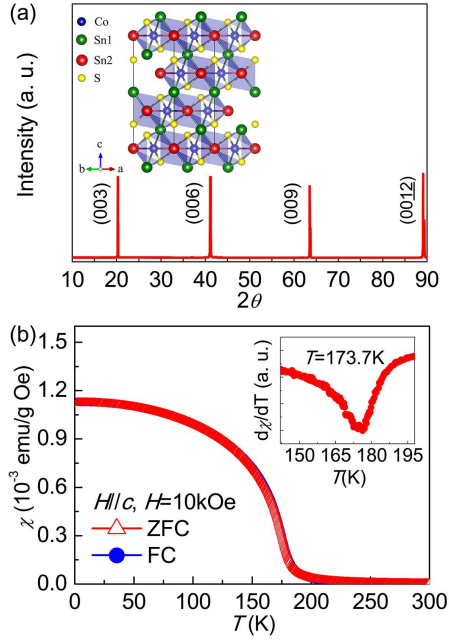}} \vspace*{-0.3cm}
\caption{(a) XRD pattern of a Co$_{3}$Sn$_{2}$S$_{2}$ single crystal. Inset: crystal structure of Co$_{3}$Sn$_{2}$S$_{2}$ with Co, Sn1, Sn2 and S atoms representing by the blue, green, red and yellow balls, respectively. (b) Temperature dependence of magnetic susceptibility $\chi(T)$ with ZFC and FC modes at $H=$ 10 kOe from 2 K to 300 K. Inset: Temperature dependence of $d\chi/dT$.}
\end{figure}

The crystal structure of Co$_{3}$Sn$_{2}$S$_{2}$ is shown in the inset of Fig. 1(a). One Co atom is coordinated with four Sn atoms (Sn2 site) and two S atoms, forming a tetragonal bipyramid (distorted octahedron). These octahedra connect each other along $ab$-plane by face-sharing and along $c$-axis by corner-sharing. The slabs of CoSn$_{4}$S$_{2}$ octahedra stack in a hexagonal packing (A-B-C fashion) along $c$-axis. Another Sn atoms (Sn1 site) are situated in between CoSn$_{4}$S$_{2}$ slabs, interconnecting them. The XRD pattern of a single crystal can be indexed by the indices of (00l) lattice planes (Fig. 1(a)), indicating that the crystal surface is normal to the $c$-axis with the plate-shaped surface parallel to the $ab$ plane. Fig. 1(b) shows the temperature dependence of the magnetic susceptibility $\chi(T)$ with the zero-field-cooling (ZFC) and field-cooling (FC) modes at an applied field of 10 kOe, suggesting this system has a ferromagnetic phase transition with $T_{C}\sim$ 173.7 K (defined as the temperature corresponding to the peak of $d\chi/dT$ curve (inset of Fig. 1(b)). It is consistent with previous reports.\cite{WangQ,Umetani}

\begin{figure}[tbp]
\centerline{\includegraphics[scale=0.3]{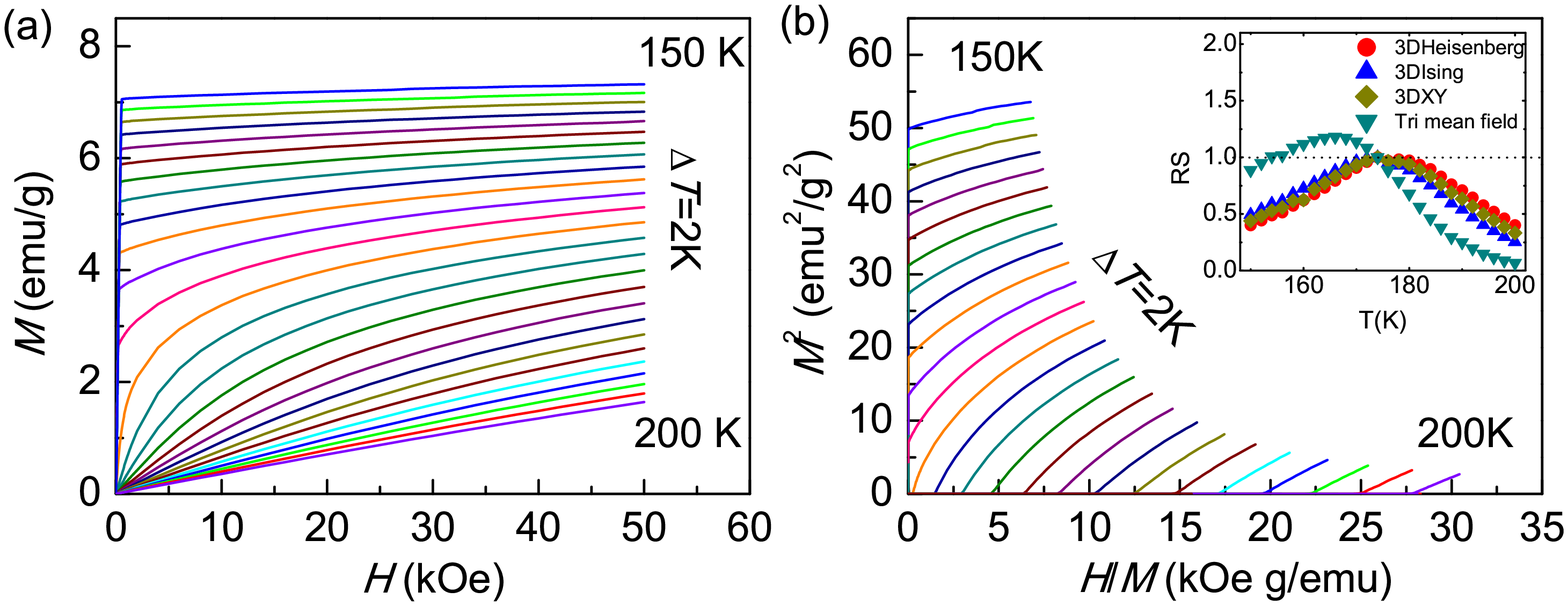}} \vspace*{0.3cm}
\caption{(a) The initial isothermal magnetization around $T_{c}$. (b) Arrott plot with mean field model. Inset: Relative slope (RS) as a function of temperature. }
\end{figure}

Fig. 2(a) displays the isothermal magnetization curves $M(H)$ of Co$_{3}$Sn$_{2}$S$_{2}$ measured in the temperature range from 150 K to 200 K with temperature steps of 2 K. The shapes of $M(H)$ curves is typical for ferromagnet. The curves below $T_{C}$ rise sharply at low-field region and saturate at high-field region. In contrast, when $T>T_{C}$, the curves gradually become linear. For determining the type of magnetic phase transition, we firstly plotted the curves of $M^{2}$ vs. $H/M$, known as the standard Arrott plot (Fig. 2(b)). Based on the criterion,\cite{Banerjee} the magnetic transition order can be determined from the slope of the standard Arrott plot. The negative slope indicates the first-order transition while the positive slope indicates the second-order one. Apparently, the positive slope of the $M^{2}$ vs $H/M$ indicates that Co$_{3}$Sn$_{2}$S$_{2}$ undergoes a second-order PM-FM phase transition in the vicinity of $T_{C}$. However, in the long-range Landau mean-field theory, the curves of $M^{2}$ vs. $H/M$ at various $T$ should form a progression of parallel straight lines in the high-field region with the line at $T = T_{C}$ passing through the origin. But the Arrott plot presented in Fig. 2(b) are not linear in the high-field region, it implies that the mean-field theory can not describe the critical behavior of Co$_{3}$Sn$_{2}$S$_{2}$. Thus, the $M(H)$ results are analyzed according to the Arrot-Noaks equation of state $(H/M)^{1/\gamma}=a\varepsilon+bM^{1/\beta}$, where $\varepsilon=(T-T_{C})/T_{C}$ is the reduced temperature, $a$ and $b$ are coefficients, and the critical exponents $\beta$, $\gamma$ are respectively associated with the spontaneous magnetization $M_{s}(T) = \lim\limits_{H\rightarrow 0}H$ and inverse initial susceptibility $\chi_{0}^{-1}(T) = \lim\limits_{H\rightarrow 0}(H/M)$,\cite{Fisher}

\begin{equation}
M_{S}(T,0)=M_{0}(-\varepsilon)^{\beta}, \varepsilon<0, T<T_{C}
\end{equation}

\begin{equation}
\chi_{0}^{-1}(T)=(h_{0}/M_{0})\varepsilon^{\gamma}, \varepsilon>0, T>T_{C}
\end{equation}

where $M_{0}$, and $h_{0}/M_{0}$ are the critical amplitudes. Correct values of $\beta$ and $\gamma$ will lead to a bunch of parallel straight lines of $M^{1/\beta}$ vs. $(H/M)^{1/\gamma}$ at various temperatures. Moreover, the line at $T = T_{c}$ should pass through the origin. Using the critical exponents of 3D Heisenberg model ($\beta=0.365$ and $\gamma=1.336$), 3D Ising model ($\beta=0.325$ and $\gamma=1.24$), 3D XY model ($\beta=0.345$ and $\gamma=1.316$), and tricritical mean-field model($\beta=0.25$ and $\gamma=1$), the relation between $(M_{s})^{1/\beta}$ and $(H/M)^{1/\gamma}$ can be established (known as the modified Arrott plot (MAP)). We further define the relative slopes RS$=S(T)/S(T_{C})$, where $S(T)$ is the slope of $(M_{s})^{1/\beta}$ vs. $(H/M)^{1/\gamma}$ at $T$. This parameter is a useful method to detect the validity of the MAP.\cite{Mahjoub,Mnassri} In an ideal model, the proper choice critical exponents will produce a MAP with a series of parallel straight lines in the high-field region, so the value of RS should equal to 1. As shown in the inset of Fig. 2(b), the deviation of tricritical mean-field model is largest and other models also can not lead to the parallel linear relation between $(M_{s})^{1/\beta}$ and $(H/M)^{1/\gamma}$. It clearly indicates that the critical behavior of Co$_{3}$Sn$_{2}$S$_{2}$ could not to be described by these universality classes. Thus, the iteration method \cite{Phan} with the initial trail values $\beta$ (= 0.325) and $\gamma$ (= 1.24) is used to extract the correct critical exponents. Briefly speaking, linear extrapolation of high-field straight line portion of the isotherm gives the value of $M_{s}(T)$ and $\chi_{0}^{-1}(T)$ as an intercepts on the $(M_{s})^{1/\beta}$ and $(H/M)^{1/\gamma}$ axes, respectively. Use these values of $M_{s}(T)$ and $\chi_{0}^{-1}(T)$, we can get a new set of $\beta$ and $\gamma$ values by Eqs.(1) and (2) and then construct a new MAP. This procedure was repeated until the values stable. The converging values of $\beta$ and $\gamma$ obtained from the good fits of $M_{s}(T)$ and $\chi_{0}^{-1}(T)$ as a function of temperature are 0.356(3) and 1.26(2)(Fig. 3(a)), which are different from those exponents in above models. The fitted $T_{C}$s are also close to that derived from the $M(T)$ curve (inset of Fig. 1(b)). On the other hand, as shown in Fig. 3(b), it is obvious that all curves in MAP are straight lines and parallel to each other with the line at $T_{C}$ just passing through the origin.

\begin{figure}[tbp]
\centerline{\includegraphics[scale=0.45]{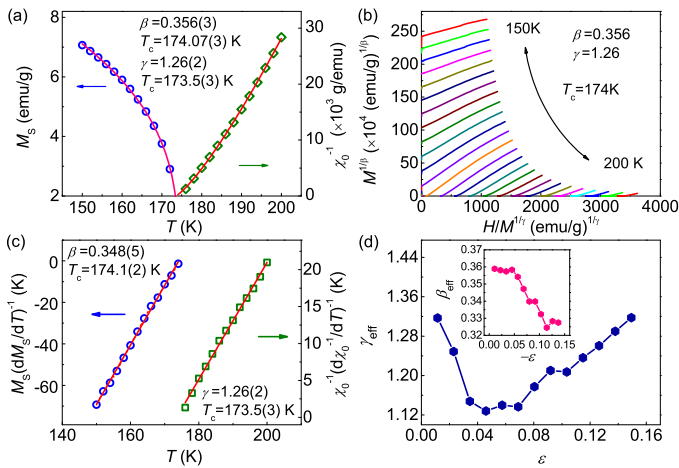}} \vspace*{0.3cm}
\caption{(a) Temperature dependence of the spontaneous magnetization $M_{s}$ (left) and the inverse initial susceptibility $\chi_{0}^{-1}$ (right) with solid fitting curves. (b) Modified Arrott plot isotherms $(H/M)^{1/\gamma}$ vs. $(M_{s})^{1/\beta}$ with $\beta=$ 0.356 and $\gamma=$ 1.26. (c) The Kouvel-Fisher plot for the temperature dependence of the spontaneous magnetization $M_{s}$ and (left) and the inverse initial susceptibility $\chi_{0}^{-1}$ (right) with solid fitting curves. (d) Effective exponents $\beta_{eff}$ (inset) and $\gamma_{eff}$ as a function of reduced temperature $\varepsilon $.}
\end{figure}

Alternatively, the critical exponents can also be determined by the Kouvel-Fisher (KF) method,\cite{Fisher}

\begin{equation}
\frac{M_{s}(T)}{dM_{s}(T)/dT}= \frac{(T-T_{C})}{\beta}
\end{equation}

\begin{equation}
\frac{\chi_{0}^{-1}(T)}{d\chi_{0}^{-1}(T)/dT}=\frac{(T-T_{C})}{\gamma}
\end{equation}

According to Eqs. (3) and (4), plots of $M_{s}(T)/dM_{s}(T)/dT$ and $\chi_{0}^{-1}(T)/d\chi_{0}^{-1}(T)/dT$ against $T$ yield linear curves with slopes of $1/\beta$ and $1/\gamma$, respectively, on which the intercepts on the abscissa axis are corresponding to $T_{C}$s. The critical exponents obtained from the KF method are $\beta=$ 0.358(5) with $T_{c}=$ 174.4(2) K and $\gamma=$ 1.311(2) with $T_{C}=$ 173.7(1) K (Fig. 3(c)), consistent with those derived from the MAP.
On the other hand, temperature-dependent effective exponents $\beta_{\rm eff}$ and $\gamma_{\rm eff}$ in the asymptotic regime ($\varepsilon\rightarrow$ 0) could reflect the existence of several competing couplings and/or disorder.\cite{Perumal2} Both of them are defined as,

\begin{equation}
\beta_{\rm eff}(\varepsilon) = \frac{d[\ln M_{s}(\varepsilon)]}{d(\ln|\varepsilon|)};
\gamma_{\rm eff}(\varepsilon) = \frac{d[\ln \chi_{0}^{-1}(\varepsilon)]}{d(\ln\varepsilon)}
\end{equation}

The $\beta_{\rm eff}$ and $\gamma_{\rm eff}$ as a function of reduced temperature $\varepsilon$ are plot in Fig. 3(d). The $\beta_{\rm eff}$ almost changes monotonically with $\varepsilon$, while the $\gamma_{\rm eff}$ shows a nonmonotonic evolution with $\varepsilon$, indicating that the $\beta_{\rm eff}$ and $\gamma_{\rm eff}$ do not match any predicted university class even in the asymptotic region. Similar phenomenon was reported in partially frustrated alloys, where the nonmonotonic changes were attributed to existence of magnetic disorders.\cite{Perumal,AKPramanik} But because the critical exponents of a magnet should be independent of the microscopic details of the system due to the divergence of the correlation length in the vicinity of the second-order phase transition, and the effective exponents are nonuniversal properties, the static critical exponents obtained from the MAP and KF method should be intrinsic.

\begin{figure}[tbp]
\centerline{\includegraphics[scale=0.22]{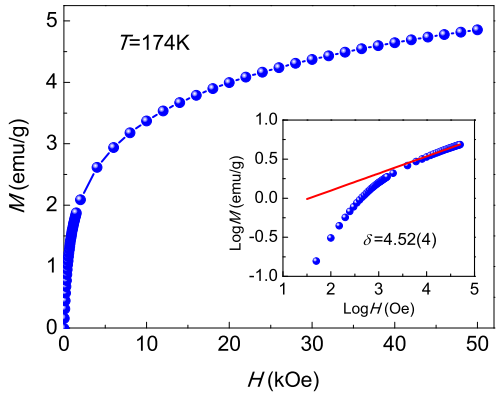}} \vspace*{0.3cm}
\caption{Isothermal M-H plot at $T_{c}$= 174 K. Inset: same plot in log-log scale with the solid linear fitting.}
\end{figure}

The critical exponent $\delta$ is associated with the magnetization isotherm at $T_{C}$ by the relationship $M=DH^{1/\delta}$, where $D$ is the critical amplitudes.\cite{Fisher} It can be obtained from the fitting of high-field slope of the $\log(M)$ vs. $\log(H)$ isotherms plot at $T_{C}$. Fig. 4 depicts the isotherm at 174 K, and the inset displays the plot on the log-log scale. The linear part in the high-field region gives the $\delta$ value of 4.52(4). In addition, the critical exponents from the static scaling analysis are related to the Widom scaling relation, $\delta=1+\gamma/\beta$.\cite{widom} Using the estimated values of $\beta$ and $\gamma$ from the MAP and the KF method yield $\delta=$ 4.662 and 4.631, respectively, close to above value of $\delta$. This, therefore, confirms that the critical exponents obtained by the magnetization data are reliable and in agreement with the scaling hypothesis. Furthermore,there is an important criterion for the critical regime based on the prediction of scaling hypothesis,\cite{widom}

\begin{equation}
M(H,\epsilon)=\epsilon^{\beta}f_{\pm}(H/\epsilon^{\beta+\gamma})
\end{equation}

where the functions $f_{+}$ for $T>T_{C}$ and $f_{-}$ for $T<T_{C}$, respectively, are the regular functions. Based on the reasonable values of $\beta$ and $\gamma$, the $M|\varepsilon|^{-\beta}$ as a function of $H|\varepsilon|^{-\beta+\gamma}$ plots in the critical region should fall onto two universal curves: one above $T_{C}$ and another below $T_{C}$. The plot of $M|\varepsilon|^{-\beta}$ against $H|\varepsilon|^{-\beta+\gamma}$ close to the critical region is selectively depicted in Fig. 5 with the values of $\beta$ and $\gamma$ obtained from the KF method. Logarithmic scaled $m$ vs. scaled $h$ has been plotted in the inset for better clarity. It can be clearly seen that all of data fall onto two universal curves, verifying the critical behavior over the measured temperature range and hence ensuring the accuracy of critical exponents.

\begin{figure}[tbp]
\centerline{\includegraphics[scale=0.23]{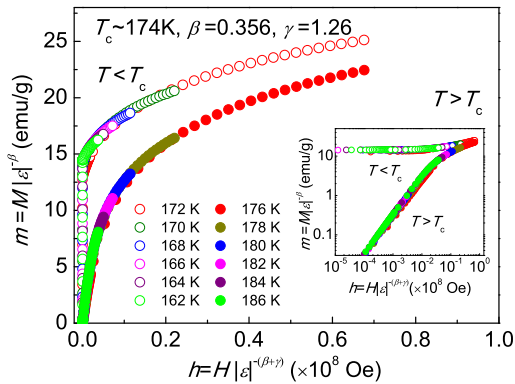}} \vspace*{-0.3cm}
\caption{Scaling plots below and above $T_{c}$ using $\beta$ and $\gamma$ determined from the KF method. Inset: same plot in log-log scale.}
\end{figure}

According to the normalization group analysis of exchange-interaction systems the universality class of the magnetic phase transition depends on the exchange distance $J(r)$. The long-range exchange interaction decays as $J(r)\sim r^{-(d+\sigma)}$ when the short-range exchange interaction decays as $J(r)\sim e^{-r/b}$, where $\sigma$ is the range of the interaction, $r$ is the distance, $d$ is the dimensionality, and $b$ is the spatial scaling factor.\cite{SFFischer} Both of these spin interactions depends on the parameter of $\sigma$, that can be calculated by the equation,\cite{MEFisher,AKPramanik,SFFischer}

\begin{multline}
\gamma=1+\frac{4}{d}\frac{n+2}{n+8}\Delta \sigma+ \\
\frac{8(n+2)(n-4)}{d^{2}(n+8)^{2}}\times[1+\frac{2G(\frac{d}{2})(7n+20)}{(n-4)(n+8)}] \Delta \sigma^{2}
\end{multline}

where $n$ is the spin dimensionality, $\Delta\sigma=\sigma-\frac{d}{2}$, $G(\frac{d}{2})=3-\frac{1}{4}(\frac{d}{2})^{2}$. Based on this equation, $\sigma<$ 2 indicates long range of spin interaction, while $\sigma>$ 2 indicates the short range of spin interaction. Since the experimental value of $\gamma$ is 1.26, with any combination of $\{d:n\}$ the values of $\sigma$ are always smaller than 2, indicating that the theoretically predicted 3D short-range type could not support the exchange interactions in Co$_{3}$Sn$_{2}$S$_{2}$ spin systems. To the present case, the $\sigma$ value should be determined by the isotropic long-range exchange interactions of the form $J(r)\sim r^{-(d+\sigma)}$. The remaining exponents can be calculated from the following expressions: $\nu=\gamma/\sigma$, $\alpha=2-\nu d$, $\beta=(2-\alpha-\gamma)/2$, $\delta=1+\gamma/\beta$. Using the experimental value of $\gamma$, other critical exponents for different sets of $\{d:n\}$ are calculated and listed in Table 1 for comparison. The calculated exponents do not match well with any of the 3D models and the presently determined values of $\beta=0.356(3)$ and $\delta=4.52(4)$ match those calculated values for a $d=$ 2 and $n=$ 1 ferromagnet with a long-range interaction between the spins decaying with distance as $J(r)\sim r^{-3.28}$. It implies that the asymptotic critical behavior of Co$_{3}$Sn$_{2}$S$_{2}$ corresponds to spin system with space and spin dimensionality of $d=$ 2 and $n=$ 1. Similar phenomenon has also been observed in various other itinerant magnetic systems, such as Pr$_{0.5}$Sr$_{0.5}$MnO$_{3}$, Cr-Fe alloy, Y$_{2}$Ni$_{7}$ systems.\cite{AKPramanik,SFFischer,ABhattacharyya} Moreover, it is also consistent with layered structure of Co$_{3}$Sn$_{2}$S$_{2}$ in which the easy-axis is along the $c$ axis with small saturated moment ($\sim$ 0.3 $\mu_{B}/$Co).\cite{WSchnelle}

\begin{table}[tbp]\centering
\caption{Comparison of Critical exponents of Co$_{3}$Sn$_{2}$S$_{2}$ with different sets of $\{d:n\}$.}
\begin{tabular}{lllllll}
\hline\hline

Critical              & $d=3$     & $d=3$     & $d=3$        & $d=2$     & $d=2$     & $d=2$  \\
exponents             & $n=1$     & $n=2$     & $n=3$        & $n=1$     & $n=2$     & $n=3$  \\ \hline

$\sigma$              & 1.9276   & 1.8609     & 1.8215       & 1.2758    & 1.2336    &  1.2080  \\
$\alpha$              & 0.0389   & -0.0312    &-0.0752       & 0.0248    &-0.0425    & -0.0848  \\
$\beta$               & 0.3505   & 0.3856     & 0.4076       & 0.3575    & 0.3913    &  0.4124  \\
$\delta$              & 4.5947   & 4.2674     & 4.0911       & 4.5238    & 4.2201    &  4.0553  \\

\end{tabular}
\label{TableKey}
\end{table}

\section{Conclusion}

In summary, the critical exponents are comprehensively investigated in the vicinity of the PM-FM transition in HMFM Co$_{3}$Sn$_{2}$S$_{2}$. This transition is found to be a second-order phase transition. The critical exponents obtained from various techniques are consistent with each other, and the experimentally obtained isothermal magnetization curves measured at different temperature collapses into two universal branches below and above $T_{C}$. It confirms that the obtained exponents are intrinsic feature of transition in Co$_{3}$Sn$_{2}$S$_{2}$. Further analysis suggests that the exchange interaction is long-range, decaying with distance as $J(r)\sim1/r^{-3.28}$, and the space and spin dimensionalities are $d=2$ and $n=1$, respectively. These critical behaviors agree well with the weak itinerant character of layered HMFM Co$_{3}$Sn$_{2}$S$_{2}$.

\section{Acknowledgments}

This work was supported by the Ministry of Science and Technology of China (No. 2016YFA0300504); the National Natural Science Foundation of China (No. 11574394, 11774423); the Research Funds of Renmin University of China (RUC) (No. 15XNLF06, 15XNLQ07); the Fundamental Research Funds for the Central Universities(No. 2017RC20, 2017RC02); and the Research Innovation Fund for College Students of Beijing University of Posts and Telecommunications.

\end{document}